# Si and N - vacancy color centers in discrete diamond nanoparticles: Raman and fluorescence spectroscopic studies

K. Ganesan,[a,*] P. K. Ajikumar,[a] S. Ilango,[a] G. Mangamma [a] and S. Dhara [a]

[a] *Surface and Nanoscience Division, Materials Science Group, Indira Gandhi Centre for Atomic Research, HBNI, Kalpakkam - 603102, India.*

*\* Corresponding author: kganesan@igcar.gov.in*

## Abstract

The present study reports on an innovative method to prepare discrete diamond nanoparticles or nanodiamonds (NDs) with high structural and optical quality through top-down approach by controlled oxidation of pre-synthesized nanocrystalline diamond (NCD) film. These NDs are studied for their structural and optical properties using atomic force microscopy (AFM), Raman and fluorescence (FL) spectroscopy. While AFM analysis confirms uniform distribution of discrete NDs with different sizes varying from a few tens of nanometers to about a micron, spectroscopic investigations reveal the presence of impurity - vacancy related color centers exhibiting FL at 637 and 738 nm as a function of particle size. In addition, an intense emission originating from vacancy centers associated with N and Si (SiV⁻) is observed for all NDs at the temperature close to liquid nitrogen. A detailed spectral analysis is carried out on the structural defects in these NDs. The full width at half maximum of diamond Raman band ( ~ 1332 $cm^{-1}$) is found to be as narrow as 1.5 $cm^{-1}$ which reveals the superior structural quality of these NDs. Further, mapping of diamond Raman and FL spectra of SiV⁻ confirm the uniform distribution of NDs throughout the substrate. The narrow line widths and a minimal shift in peak positions of Raman and FL spectra endorse this aspect.

**Introduction**

In recent years, diamond nanoparticles or nanodiamonds (NDs) have attracted much attention of researchers due to their unique structural, chemical, biological, mechanical, and optical properties.[1,2,3,4,5,6] Owing to these exotic characteristics, NDs are potential candidate in many engineering applications related to nano- and bio-technology.[2,5,6,7,8,9,10] Especially, fluorescence (FL) characteristics of NDs that originate from optical color centers, impurities and point defects have emerged as the current field of intense research for their ability as a single photon emitter in quantum information technology. [1-10]

Among various color centers in diamond, the N-vacancy (NV) color center is the most popular due to its excellent optical characteristics and hence is explored extensively. The NV center forms when a substitutional N pairs with adjacent vacancy in diamond lattice. It emits single photon at 575 and 637 nm when its electronic states of the point defect are $NV^0$ and $NV^-$ respectively. $NV^-$ emission has remarkable photo-stability, long fluorescence life time (~ 15 ns) and high quantum yield ~ 70% and thus, finds tremendous interest in room temperature (RT) nano-photonics, nanoscale imaging and sensing of bio-molecueles with high spatial resolution, and optical markers in drug delivery system. [4,6,7,10] However, poor spectral quality with a very broad phonon side bands with only < 4% of its FL into zero phonon line (ZPL) poses a serious difficulty and limit the applications of $NV^-$ centers for quantum information processing and applications.[6]

On the other hand, the negatively charged Si- vacancy ($SiV^-$) centers in diamond emerges as an alternative to $NV^-$ with its superior spectral properties.[11] In $SiV^-$, Si atom occupies position at halfway between two vacant carbon sites and forms a split-vacancy structure along the <111> axes of the diamond lattice.[11] Thus, the $SiV^-$ center has doubly split ground and excited electronic states which in turn result in emission of quartet lines observed at low temperatures.[12] The $SiV^-$ emits photons at 738 nm and has ~ 80% of emission into ZPL, narrow band width (< 1 nm), low FL background and also short FL life time of ~ 1 ns.[13] Recently, ultrafast all-optical coherent control of single $SiV^-$ color centers and coherent control of the $SiV^-$ spins in diamond are demonstrated.[5,13] Thus, the $SiV^-$ centers are considered as a potential alternative for single photon emitter in nano-photonics and the present report emphasis the merit of the case in NDs.

Point defects and impurities significantly alter materials properties and the deviation become significant as dimension of materials approach nanometer-regimes. In particular, nature of FL of ND depends on host matrix, such as dimension, morphology, defect and impurity concentration, and surface passivation.[14] A great deal of experimental and theoretical researches are currently being pursued in order to tune such emission characteristics of the NDs. Reports from literature show that the luminescent activity of $NV^-$ strongly depends on size, which has to be > 40 nm and N concentration to obtain the maximum intensity.[14,15] However, on the other hand the $SiV^-$ is stable and luminescent down to the lateral size of ~ 2 nm as proved by experiments and modeling.[2,16] Further, FL emission from $SiV^-$ enhances amidst adequate concentration of $NV^-$ in diamond lattice.[17] These observations give the impetus to the current work to address these issues based on atomic force microscopy (AFM), Raman and FL spectroscopy. Also, in this study, Raman and FL spectra are used as benchmark standard for assessing the structural quality of the NDs.

The NDs are generally described as the $sp^3$ rich tetrahedral bonded carbon material with dimensions spanning from a few nanometers to few hundreds of nanometers. Such superior optical properties demands synthesis of NDs with high structural quality. Among the various methods for NDs preparation, the detonation of TNT/RDX in oxygen deficient environment is the most popular technique and having the capacity to make large scale production with good control over particle dimension.3  Though the detonation NDs have desirable characteristics, there are a few drawbacks such as large defect concentration with $sp^2$ rich carbon layers covering the surface. The presence of $sp^2$ bond quenches the luminescence and hence, the sample requires surface passivation. [4] Further, detonation NDs (DNDs) are generally dispersed in solvents to prevent agglomeration which hinders basic studies on individual NDs. The NDs can also be produced by chemical vapor deposition (CVD),[18] high energy ball milling of diamond microcrystals,[19] laser ablation,[20] and irradiation of graphitic carbon.[21,22]  A setback, in general, in these processes are the unintentional termination of hydrogen (C-H) bonding, large surface defects and formation of $sp^2$-rich carbon, which are all known to quench the luminescence in diamond.[23]  In the current study, this problem is circumvented with C=O bonding through thermal oxidation which enhances the luminescent emission of the NDs.[23,24] Moreover, oxidation process also helps in reducing grain size by removing surface carbon layers, which eventually lead to formation of NDs. The structural quality of these NDs is extremely good due to the removal of non-diamond carbon phases and other organic contaminants, resulting in reduction of background luminescence.[25]

Despite on extensive and numerous reports available on preparation of NDs and their applications in different fields, a comprehensive understanding on the fundamental properties, formation of color centers, and influence of point defects on florescence in NDs necessitate further investigations. Herein, there are several reports exist on the oxidation behavior of diamond films with nano- or micro-crystallites, and DNDs, [26,27,28,29] but we could find hardly any on thermal oxidation to produce discrete NDs with uniform particle distribution having high structural quality with < 1.5 $cm^{-1}$ full width at half maximum (FWHM) of Raman band.

In this study, we synthesized discrete NDs over Si with a homogeneous distribution of varying lateral sizes by controlled thermal oxidation of pre-synthesized nanocrystalline diamond (NCD) grown using hot filament CVD. Having confirmed the morphological features using AFM, an extensive defect analysis and FL characteristics of these NDs are studied using micro-Raman spectrometer. Further, the size dependent Raman and luminescence characteristics of the NDs are carried out with a detailed analysis.

**Experimental methods**

**Synthesis**

In this work, NDs are synthesized by two step process. The first step involves synthesis of NCD film which acts as the starting material. The second step of this process is to oxidize the film to obtain discrete NDs. To this end, we have synthesized NCD films by a custom designed hot filament chemical vapor deposition and details pertaining to the deposition system are reported elsewhere.[30] Mechanically scratched Si(111) wafer with micron level diamond paste was used as the substrate. $CH_4$ and $H_2$ were used as feedstock gases in the ratio of 3:100 whereas substrate and filament temperature were at ~ 800 and 2000 ºC, respectively. The operating pressure in the deposition chamber was maintained at ~ 20 mbar throughout the synthesis using a throttle valve. A uniform NCD films with very good adhesion was obtained over the entire area of the Si wafer with a thickness of ~ 2 microns. This diamond film serve the purpose as a starting material for the synthesis of discrete NDs by thermal oxidization. For oxidization, the pre-synthesized NCD film was kept inside a horizontal cylindrical furnace under ambient atmosphere. Then, the furnace temperature was raised to 700 $^0$C with a heating ramp of at 10 $^0$C per minute and dwell for 15 min at the same temperature, with a help of PID controller. Subsequently, the furnace was cooled down to room temperature at 15 $^0$C per minute.

**Atomic force microscopy**

Surface topography of the NDs were measured using an atomic force microscope (NTGRA, NT-MDT, Russia). Cursory measurements on the microstructures were made with optical microscope and a detailed analysis to examine the nature of the as-grown surface and oxidized films were carried out using AFM. Based on observations made with optical microscope, surface of the film was classified into three different regions of interest (ROI) viz. (a) ROI-1, cluster free very smooth surface with no trace of particles (b) ROI-2, a very feeble to medium contrast particles and (c) ROI-3, particles with ~ 1 micron lateral size. The particle size distribution on these different regimes was measured by AFM analysis. Further, the structural and optical characteristics of these NDs were assessed through Raman and FL spectroscopic measurements as discussed in the next section.

**Raman and Fluorescence spectroscopy**

Raman and FL spectroscopic studies of NDs on the three different regions as identified by optical microscope were carried out using a micro-Raman spectrometer (Invia, Renishaw, UK) with 532 nm diode laser as the excitation source. The spectrometer was equipped with a grating monochromator (2400 / 1800 grooves/mm) and Peltier cooled charged couple device as detector. The spectral resolution of the spectrometer is 0.5 cm$^{-1}$. For a select range viz. 1000 – 2000 cm$^{-1}$ and 720 – 760 nm, Raman and FL spectra were recorded from the same location in order to evaluate the structural and optical properties of the NDs in *static mode* using a grating with 2400 grooves/mm. On the other hand, the full range of FL from 535 to 800 nm was recorded in *extended mode* using a grating with 1800 grooves/mm. Low temperature Raman and FL spectroscopic studies were also performed at 80 K using Linkam (THMS600) stage. The laser power was always kept below 2 mW unless otherwise specifically mentioned. The maximum attainable laser power is ~ 16 mW. While an objective lens of 100X magnification and numerical aperture of 0.85 (laser beam diameter ~ 1 µm) was used for RT measurements, 20X and 0.4 (laser beam diameter ~ 2 µm) combination with long focal length was used for low temperature, so as to accommodate the special sample holder. Spectra, thus acquired, were analyzed using WIRE 4.2 software.

**Results**

Figure 1a shows the topographical images of the as-grown NCD film and depicts the surface compromising diamond crystallites of large size distribution. Upon thermal oxidation, the diamond crystallites undergo disintegration in their size by CO and $CO_2$ formation. This leaves only the large particles to survive with reduction in their size as NDs. Fig.1b shows the optical micrograph of the thermally oxidized film which clearly displays the homogeneous distribution of discrete particles with a distribution in their sizes. In ROI-1, the surface appears to be very smooth with no features. However, the presence of particles may not be ruled out as particles with less than 300 nm size falls below the observable limit of the optical microscope. In ROI-2, surface contrast is a very feeble to medium suggestive of small or medium sized particles with lateral dimensions in the range 300 to 800 nm and in ROI-3, the particle size is estimated to be greater than 800 nm and sparingly scattered around as compared to medium sized particle as can be evident from the Fig.1b. These regions are further analyzed with an AFM and discussed in the next paragraph.

Figure 1c and 1d show the AFM topographic features of ROI-1 and ROI-2 respectively, which clearly displays the distribution NDs with different lateral dimensions. The lateral size distribution of these ND particles are measured using semi-automatic line profile analysis by ImageJ software. Then, the size distribution is obtained within a bin size of 10 and 40 nm in ROI-1 and 2, respectively. Subsequently, a plot was drawn with a variation of number of particles as a function of particle size and shown in Fig.1e and 1f. Then, the plots were fitted to Gaussian profile with peak as the average lateral dimension of the NDs. Similarly, the maximum height of these NDs are measured using line profile analysis of AFM software and the mean and standard deviation are estimated from these height values. While the estimated mean height and lateral dimension of these NDs are found to be ~ 14 ± 5 nm and 51± 2 nm, respectively in ROI-1; they are found to be ~ 170 ± 81 nm and 310 ± 7 nm, respectively, in ROI-2.  A closer look at the topographic features reveals that a considerable number of NDs have agglomerated together to form clusters which are perceived as larger in lateral size as compared to the height of discrete NDs particles. In the case of ROI-3, the average height and lateral particle size are about 800 nm and 1 μm respectively and their images are not shown here as they are found to be as isolated entities and very less in number density. Raman and FL spectra are taken on these three regimes independently and their representative data analysis is discussed below.

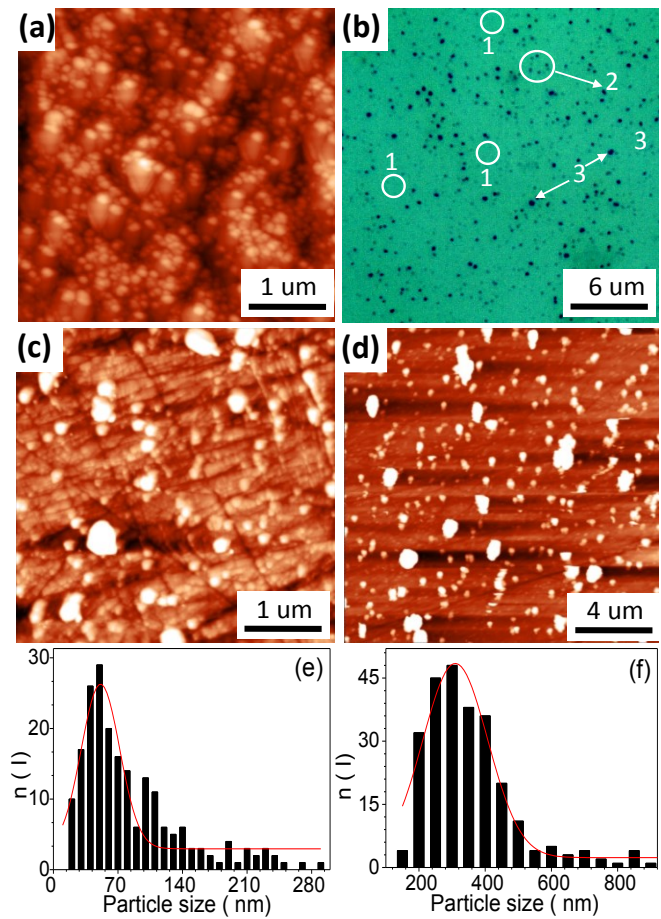

Fig.1. (a) AFM topography of the as-grown nanocrystalline diamond film. (b) Optical micrograph of the diamond particles (NDs) dispersed on Si substrate. A representative three different regions are marked on the micrograph. (c & d) AFM topography of the NDs measured over the regions 1 and 2 that are marked on the optical micrograph, respectively. (e & f) Histograms constructed from the lateral size distribution on acquired images in the regimes 1 and 2 respectively. The maximum peak to peak height variation in Fig 1a, 1c and 1d are found to be about 170, 48 and 490 nm, respectively. The z-scale is adjusted for these images appropriately in order to have maximum visibility of the NDs.

Figure 2a and 2b shows the FL and Raman spectra of as-grown NCD film and NDs over the ROI-1, 2 & 3 at room temperature. The Raman spectra of as-grown NCD film displays very broad bands at ~ 1350 and 1580 cm$^{-1}$ corresponding to D and G bands respectively, of disordered graphitic carbon as shown in inset of Fig.2b. In addition, this spectrum also consists of a broad hump at 1140 cm$^{-1}$ which is the signature of poly-acetylene structure that arises due to presence of H in the disordered carbon network.[23] The FL spectrum does exhibits the typical signature of NCD structure with two broad bands at ~ 630 and 700 nm.[31]

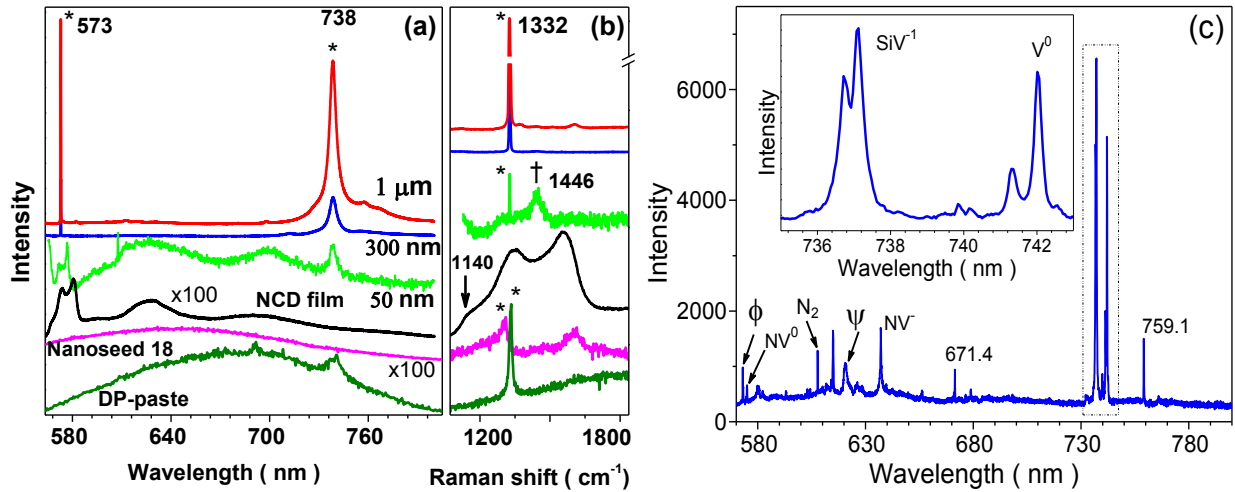

Fig.2 (a) Typical fluorescence spectra of nanodiamonds at different microstructures and lateral dimensions, (i) 50 nm (ii) 300 nm (iii) 1000 nm, (iv) pristine nanocrystalline diamond film and commercial (v) Nanoseed 18 & (vi) micron sized diamond paste, DP-Paste. (b) Raman spectra of these diamonds in the range 1075 – 1830 cm$^{-1}$. The PL spectrum of NDs with average lateral size of 50 nm and Nanoseed 18 is multiplied by 100 in order to bring in scale with other spectra. The PL spectrum of NDs with average lateral size of 50 nm and Nanoseed 18 is multiplied by 100 in order to bring in scale with other spectra. The spectra are stacked up vertically for clarity without removing background. (c) FL spectrum of nanodiamonds at 80 K and a magnified part of the spectrum is given as inset in Fig.2c.

The FL spectra of NDs consist of two prominent features at ~ 573 and 738 nm corresponding to Raman mode of diamond crystals (~ 1332 cm$^{-1}$) and FL emission from SiV$^-$ center, respectively. The absolute FL intensity of SiV$^-$ emission is found to be 150, 15400 and 65000 for the NDs with average lateral dimension of ~ 50, 300 and 1000 nm under identical measurement conditions. Thus, FL intensity of SiV$^-$ emission increases linearly with size of the NDs. The FL emission spectra of SiV$^-$ is best fit to a Lorentzian function for the NDs in the ROI 1, 2 and 3 respectively. Also, the FL spectrum of NDs in the ROI-1 shows two broad emission bands at ~ 630 and 700 nm as similar to as-grown NCD films. Also, FL spectra of NDs in ROI-1 exhibit additional prominent emission lines at ~ 576.8 and 571.6 nm (1446 & 1290 cm$^{-1}$) which are associated with third order Raman band of Si that was used as substrate in the present study. Further, it may be noted that SiV$^-$ emission is not observed in as-grown NCD film, which is possibly hindered by the surface hydrogenation or sp$^2$ rich carbon secondary phases.[23] Further, the quality of our NDs are compared with the commercial nanodiamonds ( Nanoseed 18, Microdant, Switzerland) and micron sized diamond paste (1 μm, DP paste P, Struers, Denmark). As shown in Fig 2a and 2b, the general features of the FL spectra of commercial nano- and micro-diamond products are in close agreement with the synthesized discrete diamond nanoparticles in our laboratory. However, Raman linewidth analysis reveal that the structural quality of our lab grown discrete nanodiamonds are superior as compared to the commercial diamond products (Fig. 2b).

Figure 2c shows the FL spectra of NDs in ROI-3 measured at 80 K. The spectrum consists of several well defined Raman and FL bands at different wavelengths. Raman bands originate from the phonon scattering process involving single or multiple phonons. Notably, the first and second order Raman bands of diamond are observed at 573 and 621 nm (~ 1332 and 2671 cm$^{-1}$), and represented with the symbols ϕ and Ψ, respectively in Fig 2b. In addition, several characteristic FL emission lines are also observed with prominence at 575, 637, 738 and 742 nm corresponding to neutral N- vacancy center (NV$^0$), NV$^-$, SiV$^-$ and neutral vacancy (V$^0$) color centers, respectively. In addition, there are a few sharp and intense bands at ~ 607.8, 614.9, 671.4 and 759.1 nm and these shall be discussed later. Further, a split in SiV$^-$ emission line is also observed at ~ 738 nm which is due to the doubly degenerate excited and ground states of SiV$^-$.[32] Nevertheless, these NDs exhibit considerable changes in their Raman and FL spectra depending on the structural defects and impurities. In order to study the variation of the defects and impurities in these NDs, a large number of Raman and FL spectra are recorded over the three different ROI. A part of the spectroscopic data are given in Fig. S1, S2, S3, S4 and S5 under supplementary information. As shown in supplementary figures (Fig. S1-S5), a significant variation in the spectra is observed due to the subtle differences among the diamond nanoparticles in terms of particle size and structural defects. Among the obtained spectra on NDs, we categorize the spectra with similar features in each ROI and those results are discussed in the subsequent paragraphs.

Fig. 3a, 3b and 3c depict a few representative FL spectra recorded on different spots over all the three ROI-1, 2 and 3 respectively, at 300 K. These spectra are further sectioned and are vertically shifted for easy viewing of weak bands. Fig. 3a shows the FL spectra of NDs recorded on five different particles in ROI-3. These spectra display strong Raman and FL bands at 1332 cm$^{-1}$ and 738.4 nm corresponding to SiV$^-$, respectively. Further, it is noticed that GR1 center, due to V$^0$ in diamond, at ~ 741.5 nm is observed on several diamond particles in ROI-3, as can be seen from the spectra S1 and S2 in Fig.3a.[9] Also, the FL emission from NV$^-$ center at 637 nm is observed on majority of diamond particles on ROI-3 as can be seen in spectra S2-S5 (Fig.3a). In addition, these diamond particles also exhibit distinct sharp Raman and FL bands at different energies between 585 and 800 nm as shown in Figs. 3a and 3b.

Figure 3b depicts the representative FL spectra recorded in the range 560 − 780 nm of the NDs in the ROI-2. As similar to ROI-3, these spectra also show a very sharp Raman and FL bands at ~ 573 ( 1332 cm$^{-1}$ ) and 738 nm respectively. In addition, emission from GR1 center is also observed rarely (*e.g.*, spectrum S10) on a few diamond particles which reveals the sparse existence of V$^0$ in NDs in ROI-2. However, FL emission from NV$^-$ center is not observed on any diamond particles in ROI-2. Also, several FL emission lines and second order Raman bands are absent as compared to the NDs in ROI-3.

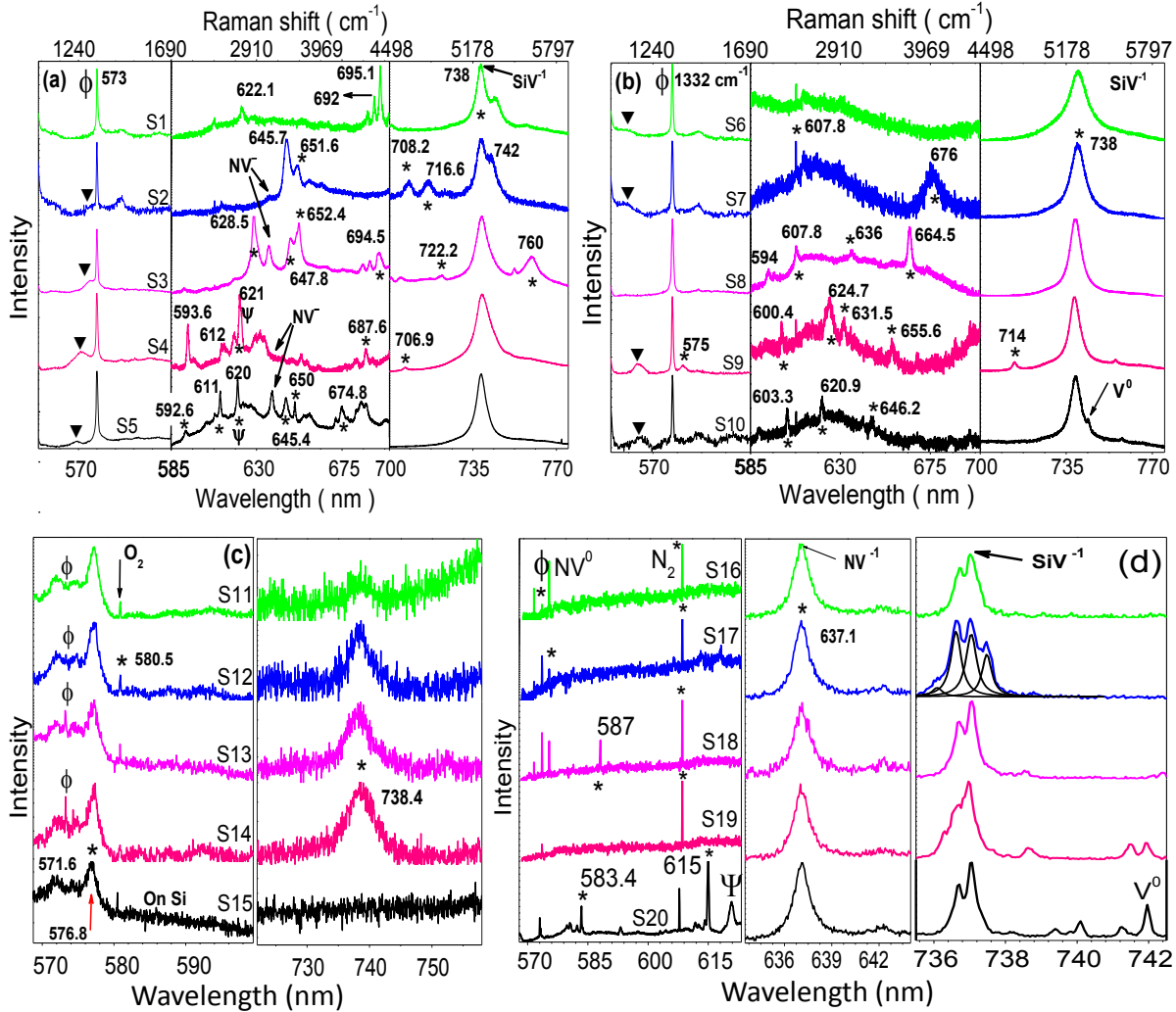

Fig.3. Representative FL spectra of nanodiamonds (NDs) recorded on different regions of interest having average lateral dimension of about (a) 1 μm, (b) 300 nm and (c) 50 nm measured at room temperature and (d) ensembles of NDs measured at 80 K. These FL spectra are further sectioned and intensity scale is adjusted in each section by linear shift for clarity in viewing all significantly weaker bands. The symbols, ϕ and Ψ are used to represent diamond first and second order overtone Raman bands (1332 and 2667 cm$^{-1}$); symbol ▼ in Fig.3a & 3b is used to represent the presence of sp$^3$ type defects, and ∗ is used to represent a peak position in the Figure.

Figure 3c exhibits Raman spectra of NDs in ROI-1 with a very sharp peak at ~1332 cm$^{-1}$ (573 nm, represented with the symbol, ϕ). However, the intensity of the diamond Raman peaks varies at different spots. The variation in Raman intensity can be correlated to the particle size and/or possible number of NDs that are present under the probing area of ~ 1 micron of the laser beam diameter. As discussed earlier, the two broad peaks at around 576.8 and 571.6 nm (1290 and
11

1446 cm$^{-1}$ ) are attributed third order Raman scattering of Si as shown as reference spectrum (S15) in Fig 3c. The sharp Raman peak at ~ 580.5 nm (1554 cm$^{-1}$) is a characteristic signature of atmospheric O$_2$ molecules. Further, FL spectra recorded at the same location exhibit emission at ~ 738 nm corresponding to the SiV$^-$ and such emission is observed even in the absence of Raman signal pertaining to the diamond. This ensures the strong luminescence from SiV$^-$ centers in all NDs. Note that, in ROI-1, the Raman and FL spectra are recorded at 8 mW in order to obtain sufficient intensity.

Figure 3d shows the representative Raman and FL spectra that are recorded in the ROI- 1, 2 and 3 at 80 K. The spectra from S16 to S20 in Fig. 3d are arranged with approximately increasing dimensions of NDs. Since the objective lens was 20X for 80 K measurements, the diameter of the laser probe becomes ~ 2 μm which restricts the independent spectroscopic investigations of NDs in ROI-1 & 2 due to overlapping. Additional Raman and luminescence bands emerge, as the lateral dimension of ND increases toward ROI-3, as shown in Fig.3d. In contrast to RT measurements, the FL from NV$^0$ and NV$^-$ centers are also observed almost in all the NDs, irrespective of the particle size (Fig.3d). The FWHM of the total SiV$^-$ band is very narrow and also SiV$^-$ emission exhibits multiple emission lines at about 738 nm. For example, the spectrum S17 (Fig.3d) is deconvoluted into four prominent lines and shows very characteristic of SiV$^-$ defects in diamond at low temperatures.[32] Other spectra in Fig.3d exhibit doublet or triplet structures and further reduction in temperature below liquid nitrogen may be required to exhibit quartet fine structures.

The structural stability and integrity of color centers in diamond lattice under high power laser irradiation are important parameters for ND based photon emitters. These properties are evaluated on the NDs in ROI-1, 2 and 3 using Raman and FL spectroscopic measurements as a function of laser power varying from ~ 0.08 to 16 mW. Fig.4 depicts the Raman and SiV$^-$ FL spectra obtained from a single ND particle in ROI- 2 and 3 with different laser power at RT. The extracted parameters of the best fit from Raman and FL spectra using Lorentzian line shape analysis are given in Table 1. The diamond Raman band intensity (area under the curve, I$_A$ ) monotonically increases with laser power upto 8 mW on both the ROI - 2 and 3 as shown in Table 1. In addition, the NDs on ROI-2 are stable even at the laser power of 16 mW. On the other hand, the NDs on ROI-3 are not stable at 16 mW and diamond Raman band undergoes a red shift when laser power goes higher than 1.6 mW as shown in Fig 4c. The Raman band red shift is attributed to laser induced temperature effects on the structural properties of diamond.[33]

The FWHM of the corresponding diamond Raman band also increases with laser power as can be seen from Table 1. In addition, FL intensity, $I_A$, of SiV¯ increases with laser power as shown in Table 1. However, the peak intensity decreases drastically at > 8 mW power. The reduction in peak intensity with increasing laser power substantiates the occurrence of laser induced thermal effect which essentially increases the non-radiative rate in these NDs at high temperatures.[8] In addition, the reduction of emission intensity can also happen due to the photo induced ionization of color centers in which the negatively charged point defects converts into neutral defects by donating an electron to diamond lattice.[34] Further, the peak position of Raman and SiV¯ bands shifts to lower wave number (Fig.4e) and higher wavelength (Fig.4f), respectively, with increasing laser power. The reason for such structural degradation can be correlated with the effect of temperature on diffusion of point defects in diamond lattice.[15] However, the structural quality of the discrete ND in ROI-2 and 3 is found to be superior at low power laser irradiation. This fact is revealed by the small values of FWHM of the Raman band with ~ 1.5 and 2.3 cm$^{-1}$ for the NDs in ROI-2 and 3, respectively (Table 1).

The NDs in ROI-1 do not undergo structural degradation even at the highest power of 16 mW and also there is no Raman and FL signature at the laser power of less than 8 mW. Further, the FWHM and peak position of Raman and SiV¯ bands remain same except an increase in intensity. Since the actual laser power received by the tiny NDs (size ~ 50 nm ) are extremely small, they are stable even at 16 mW. Moreover, owing to the high surface area of NDs for the lateral size with < 300 nm, the heat dissipation to the environment is larger as compared to that for the micron size particles. On the other hand, the NDs in ROI- 3 undergoes maximum structural degradation as they receive maximum available power density into the diamond lattice during measurements. Further, it is observed no FL emission at 637 nm from NDs in ROI- 1 and 2. Furthermore, a similar laser power dependent Raman and FL emission is measured on a few NDs in ROI-2 and 3. By and large, the spectroscopic results on NDs follow a similar behavior in each category. Additional spectra are given in Fig. S4 and S5 under supplementary information.

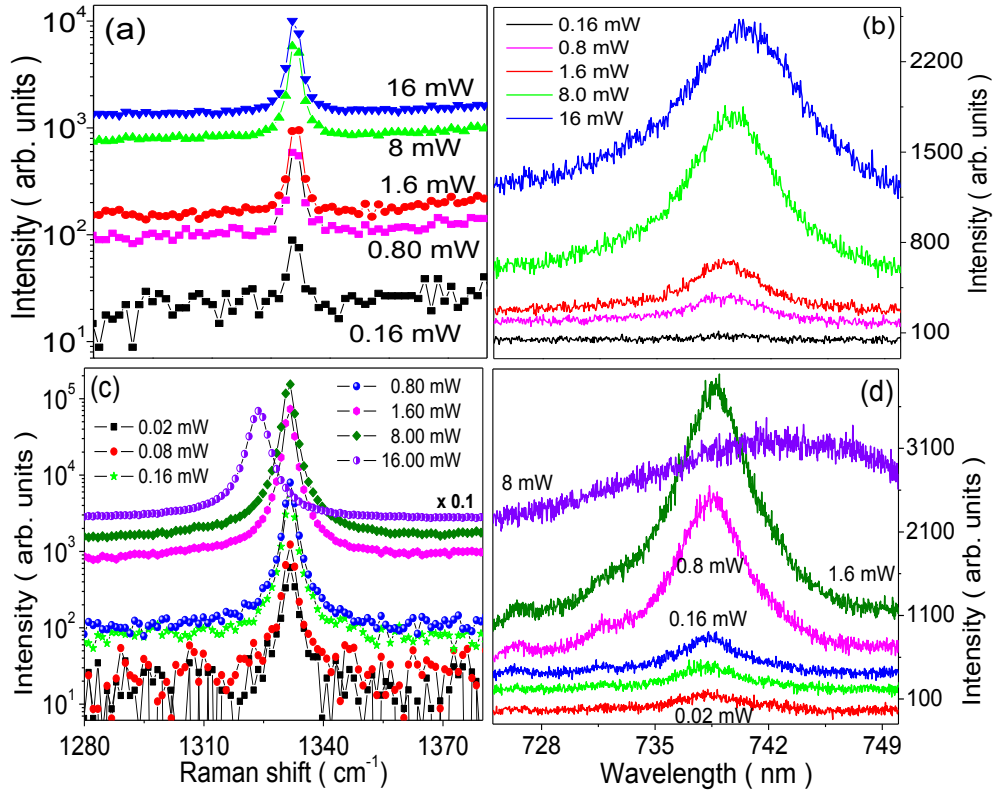

Fig.4. The intensity variation of (a) Raman and (b) FL spectra as a function of laser power on a discrete diamond nanoparticle in the region of interest - 2. Similarly, (c) Raman and (d) FL spectra as a function of laser power on a discrete diamond nanoparticle in the ROI - 3.

Table 1. The extracted parameters from best fit analysis of Raman and FL spectra from Fig. 4

| Power (mW) | Raman | | | | | | SiV⁻ emission | | | | | |
|---|---|---|---|---|---|---|---|---|---|---|---|---|
| | ROI-2 | | | ROI-3 | | | ROI-2 | | | ROI-3 | | |
| | P (cm$^{-1}$) | FWHM (cm$^{-1}$) | $I_A$ | P (cm$^{-1}$) | FWHM (cm$^{-1}$) | $I_A$ | P (cm$^{-1}$) | FWHM (cm$^{-1}$) | $I_A$ | P (cm$^{-1}$) | FWHM (cm$^{-1}$) | $I_A$ |
| 0.08 | - | - | - | 1331.6±0.02 | 2.26±0.02 | 19742±120 | 738.6±0.11 | 4.28±0.35 | 331±22 | 738.3±0.05 | 5.2±0.2 | 1631±89 |
| 0.16 | 1331.3±0.05 | 1.54±0.5 | 308±16 | 1331.6±0.01 | 2.26±0.02 | 46658±266 | 738.5±0.07 | 5.24±0.26 | 767±34 | 738.3±0.02 | 4.7±0.1 | 3130±58 |
| 0.80 | 1331.3±0.12 | 1.76±0.12 | 2365±34 | 1331.4±0.01 | 2.24±0.02 | 266936±1389 | 738.6±0.02 | 5.23±0.07 | 4070±43 | 738.4±0.01 | 5.5±0.1 | 16353±130 |
| 1.60 | 1331.4±0.01 | 1.82±0.11 | 4148±51 | 1331.3±0.01 | 2.21±0.01 | 591661±2253 | 738.6±0.01 | 5.30±0.05 | 7072±55 | 738.6±0.01 | 6.2±0.1 | 27130±184 |
| 8.00 | 1331.3±0.01 | 1.90±0.06 | 24421±51 | 1323.7±0.01 | 3.97±0.04 | 427569±2909 | 739.1±0.02 | 7.18±0.08 | 26415±271 | 742.7±0.05 | 25.8±0.6 | 58923±2486 |
| 16.00 | 1331.2±0.01 | 2.19±0.05 | 40671±354 | - | - | - | 739.6±0.04 | 9.65±0.15 | 40760±608 | - | - | - |

Based on statistical analysis of Raman spectra recorded on multiple NDs with different lateral dimension over the sample surface at RT, the FWHM of the Raman band in ROI- 1, 2 and 3 is found to be 2.2±0.5, 2.5±0.4 and 2.8±0.7 cm$^{-1}$, respectively (supplementary Fig.S1-S3). These values show an increase in FWHM as a function of mean particle size indicating possible increase in defect concentrations such as vacancies and their complexes with N and Si impurities in the diamond lattice.[35] However, the peak position of Raman band does not vary significantly among these different regions. Further, the small variation in peak position may depend on many parameters including doping of impurities like N and Si, point defects, size effects in the nanoscale regimes and more importantly laser induced thermal effects. Hence, the observed phenomenon is a manifestation of the combined effects of the above said parameters and it is very difficult to attribute for an individual parameter. After having knowledge on the spectroscopic behavior of individual NDs, their distribution is analyased through two dimensional mapping of Raman and SiV¯ FL spectra. These results are discussed in the next paragraph.

Figure 5 illustrates the two dimensional map of Raman and FL spectra of the nanodiamonds for their intensity (a,b), FHWM (c,d) and peak position (e,f) distribution. Raman and FL spectral mapping are carried out subsequently at the same location. Each image has 48 x 36 pixel data of Raman or FL spectra ( area ~ 28 x 21 μm$^2$). The intensity maps of Fig.5a and 5b are generated by picking the intensity value of Raman and FL spectra at 1332 cm$^{-1}$ and 738.4 nm, respectively on each pixels. The maximum intensity in these images is ~ 21000 and 20000 for Raman and FL emission, respectively. In order to make the low intensity particles more prominent, z-scale threshold value has been fixed at 1000. It is learnt from our point spectroscopy analysis that the intensity vary linear with particle size (Fig. 2a). Thus, the intensity variation in Fig.5a and 5b can be directly correlated with particle size. This confirms the spatially homogeneous distribution of NDs with varying sizes. This observation also supports the AFM topography for the NDs particle distribution.

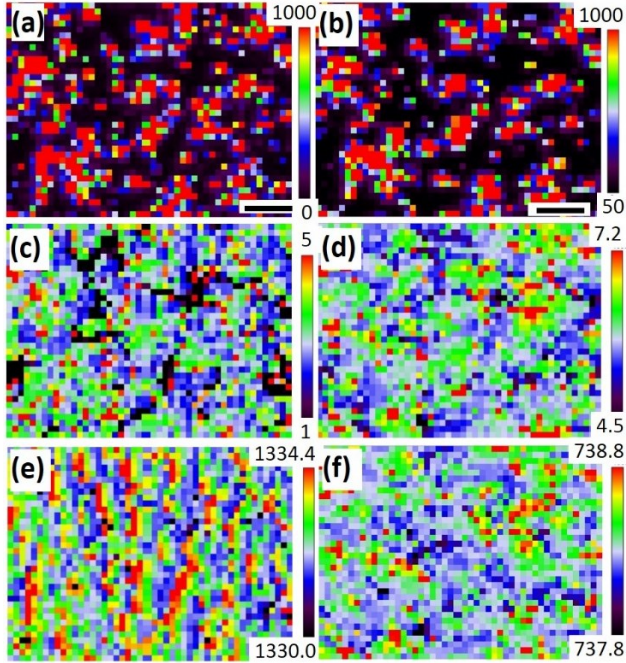

Fig.5. Two dimensional map of Raman and Si-V⁻ fluorescence spectra of nanodiamonds distribution on Si. The peak intensity distribution of (a) Raman band at 1332 cm$^{-1}$ and (b) Si-V⁻ emission at 738.4 nm. The micrographs (c) and (d) show the calculated FWHM of Raman and Si-V⁻ emission bands respectively; (e) and (f) show the peak position of Raman and Si-V⁻ emission bands respectively, as estimated from Lorentzian fit of each spectrum in the mapping. Note that the z-scale threshold values are adjusted to intermediate values to have a better visibility on the majority of the data. The scale bar and total area in the images are 5 μm and 28 x 21 μm², respectively.

Furthermore, each Raman and FL spectrum in the mapping are fitted with Lorentzian function and their calculated FWHM (Figs. 5c and 5d) and peak position (Figs. 5e and 5f) are measured. The mean FWHM and peak position of Raman spectra is ~ 2.2 ± 0.8 cm$^{-1}$ and 1332.4 ± 0.8 cm$^{-1}$ respectively, signifying presence of low defect concentration in the NDs. The SiV⁻ emission is present in almost all the NDs irrespective of the particle dimension. The mean peak position of SiV⁻ is found to be ~ 738.3 ± 0.2 nm. The mean FWHM of SiV⁻ emission is ~ 5.4 ± 0.3 nm which is comparable with the reported ones for ensembles of SiV⁻ centers in literature.[12, 32] In addition, we found a good correlation in intensities of Raman and FL mapping as computed by ImageJ software. The cross correlation coefficient between the intensity map of Raman and FL spectra is found to be 0.85. However, a similar computation made on the maps of FWHM and peak position for these corresponding images yield a poor correlation ( <0.3).

The table 2 and 3 show the fit parameters that are calculated for the spectra given in Figs.3a-3c and 3d, respectively. As can be seen from these tables, the peak position of the Raman as well as SiV⁻ bands do not vary significantly except for a minor shift ensuring the good structural homogeneities of NDs. The observed increase in FWHM of these NDs reported in Table 2 and 3, as compared to our earlier quoted values, are attributed to the longer exposure of samples to the laser irradiation while collecting the entire spectrum. Further, a comparison of peak and area intensity ratios of SiV⁻ to Raman band, *i.e.,* ($I_{SVC}/I_R$) and $A_{SVC}/A_R$, respectively reveals intensity of Raman signal is larger than that of the SiV⁻ line at RT as significant attenuation of FL intensity takes place due to surface defects. On the other hand, at 80 K, the SiV⁻ and NV⁻ centers emit significantly intense photons with very narrow line widths down to 0.60 and 1.33 nm, respectively as illustrated in Table 3. Here, the FWHM of SiV⁻ emission with 0.6 nm corresponds to the combined line widths of intense doublets at ~ 738 nm.

**Discussion**

Based on our overall observations and interpretations in the earlier paragraphs, we summarize the results as follows. By making use of top-down approach a uniform distribution of NDs are obtained by thermal oxidation of pre-synthesized nanocrsytalline diamond thin film. The FWHM of diamond Raman band is found to be 2.2±0.5, 2.5±0.4 and 2.8±0.7 cm⁻¹, respectively for NDs with increasing average lateral dimension designated as ROI - 1, 2 and 3. The broadening of Raman line width indicates that the phonon life time or the phonon mean free path decreases with increasing mean lateral dimension of NDs. Thus, the increase in line width clearly reveals the increase in defect concentration which eventually decrease the phonon life time in NDs with increasing mean particle size. This is further corroborated to the observations of additional FL emission from NV⁻ and GR1 centers with increase in lateral dimension of the NDs. Also, the estimated mean FWHM from the Raman mapping over an area of 28 x 21 μm² is about 2.2 ± 0.8 cm⁻¹ which is in good agreement to the Raman spectra obtained for ROI-1. In addition, ND's Raman line width as low as 1.5 cm⁻¹ is also observed on NDs in ROI-2 (Fig. 4b). Such a narrow bandwidth of Raman spectra confirms excellent structural quality of the NDs. In terms of comparison, these synthesized NDs are as good as defect free single crystal diamond, which has FWHM of 1.6 cm$^{-1}$.[35]

Apart from fundamental Raman mode at 1332 cm$^{-1}$, we found several additional Raman and FL bands at RT and low temperature as shown in Fig.3. The notable Raman modes at ~ 1909 (592.6), 1935 (593.6), 2199 (603.3), 2436 (611), 2446 (612.1), 2462 (612.7), 2522 (615), and 2674 (621) cm$^{-1}$ are attributed to second order Raman scattering involving two phonons in diamond lattice. The numbers in parenthesis are in "nm" unit and these bands are indicated in Fig.3 These Raman frequencies are consistent with the reported second order Raman modes by experimental and theoretical calculations.[36] In addition, the observation of Raman mode at ~ 1740 cm$^{-1}$ (587 nm) confirms the existence of C=O bonds on the surface that are formed during thermal oxidation.[37] Also, a Raman band is observed at 2330 cm$^{-1}$ (607.8 nm) in all NDs which is attributed to Raman scattering of adsorbed $N_2$ molecules on NDs. Apart from second order diamond Raman bands, there are several notable FL bands in the wavelength range 650 – 730 nm as can be seen from Fig 3. The FL bands at ~ 647 and 722 nm are mostly due to high incorporation of nitrogen in NDs.[38] The other narrow FL bands ~ 622, 624.7, 628.5, 631.5, 645.4, 646.2, 650, 651.6, 652.4, 655.6, 664.5, 676, 687.6, 692, 694.5, 695.1, 707, 708, 714 and 716.6 nm could have originated from localized point defects that are part of an extended morphological defects in individual nanodiamonds.[39] Further understanding on this aspect warrants detailed investigation.

In yet another observation, a broad Raman band with peak position in between 1250 and 1310 cm$^{-1}$ is observed as a hump with red shift to fundamental Raman mode (1332 cm$^{-1}$) on these NDs as can be seen from Fig 3a and 3b. Generally in literature, the broad hump with red shifted Raman line is attributed to irradiation induced vacancy defects and phonon confinements in diamond lattice.[40,41] However, these effects are not clearly understood so far. As reported earlier with experimental verification, we assign the broad hump that appears between at 1280 and 1320 cm$^{-1}$ to a special type of *sp³* point defect that present in interiors of grains in diamond crystal.[42] This *sp³* defects can arise during growth and do not influence on the Raman scattering either on the fundamental Raman line width or position or *sp²* bonded carbon in diamond lattice. However, these defects give rise to additional band at a lower frequency than the fundamental diamond Raman line.[42]

Table 2. The best fit parameters for the Raman & SiV⁻ center emission bands that are given in Fig. 3

| Average lateral size | Spectrum | $P_R$ (cm$^{-1}$) | FWHM (cm$^{-1}$) | $P_{SVC}$ (nm) | FWHM (nm) | $I_{SVC}/I_R$ | $A_{SVC}/A_R$ |
|---|---|---|---|---|---|---|---|
| 1 μm | S1 | 1333.1 | 4.7 | 738.3 | 4.3 | 0.59 | 0.37 |
| | S2 | 1332.2 | 4.2 | 738.3 | 4.4 | 0.88 | 0.9 |
| | S3 | 1332.3 | 4.3 | 738.8 | 6.2 | 0.32 | 0.32 |
| | S4 | 13320 | 4.7 | 739.1 | 6.7 | 0.62 | 0.6 |
| | S5 | 1332.2 | 4.7 | 738.5 | 4.7 | 1.28 | 0.88 |
| 300 nm | S6 | 1331.5 | 3.9 | 739.9 | 7.4 | 0.2 | 0.26 |
| | S7 | 1331.9 | 4.2 | 739.7 | 6.4 | 0.46 | 0.47 |
| | S8 | 1332.6 | 5.7 | 738.6 | 5.0 | 0.67 | 0.4 |
| | S9 | 1332.3 | 4.7 | 738.5 | 5.1 | 1.04 | 0.75 |
| | S10 | 1332.9 | 5.2 | 738.9 | 5.3 | 2.11 | 1.46 |
| 50 nm | S11 | - | - | 738.5 | 3.5 | - | - |
| | S12 | 1331.5 | 4.2 | 738.2 | 5.3 | 0.68 | 0.58 |
| | S13 | 1332.7 | 2.3 | 738.5 | 5.5 | 1.29 | 2.14 |
| | S14 | 1333.3 | 2 | 738.5 | 5.6 | 0.91 | 1.72 |

Table 3. The best fit parameters for the Raman, SiN⁻ and NV⁻ emission bands measured at 80 K, corresponding to the spectra in Fig 3d.

| Spectrum | $P_R$ (cm$^{-1}$) | FWHM (cm$^{-1}$) | SVC (nm) | | | NVC (nm) | | $I_{SVC}/I_R$ | $I_{NVC}/I_R$ |
|---|---|---|---|---|---|---|---|---|---|
| | | | Mean position | Comb. FWHM | FWHM/band | $P_{NVC}$ (nm) | FWHM (nm) | | |
| S16 | 1331.89 | 2.21 | 736.94 | 0.69 | 0.35 | 637.09 | 1.36 | 12.51 | 18.1 |
| S17 | 1332.20 | 2.46 | 736.81 | 1.39 | 0.34 | 637.09 | 1.16 | 3.33 | 1.50 |
| S18 | 1332.24 | 2.67 | 736.94 | 0.62 | 0.31 | 637.11 | 1.38 | 5.64 | 1.01 |
| S19 | 1332.33 | 2.23 | 736.70 | 0.95 | 0.31 | 637.08 | 1.05 | 8.40 | 0.98 |
| S20 | 1332.05 | 2.95 | 736.94 | 0.62 | 0.31 | 637.09 | 1.39 | 1.42 | 1.05 |

Further FL characteristics of these NDs are investigated through representative spectral analysis. The best fit for FL spectra of SiV⁻ and NV⁻ are found to be Lorentzian line shape. The observed line widths of SiV⁻ centers in NDs are found to vary from 3.5 to 7.4 nm with mean value of ~ 5.4 ± 0.3 nm at RT. At 80 K, the SiV⁻ emission line width further decreases with negligible phonons side bands. Also, at 80 K, FWHM of SiV⁻ is found to decrease down to 0.6 nm from 5 nm observed at RT. The FWHM of 0.6 nm is a combination of line widths of an intense doublets observed ~ 738 nm. While FWHM of NV⁻ centers is ~ 2.2 ± 0.3 at RT, it is further decreased down to 1.38 ± 0.05 nm at 80 K. These FL line widths are very narrow and suggestive of a minimal electron – phonon interactions in the NDs. Although the line width from SiV⁻ color center is found to be narrow, this value is still high when compared to single photon emitters, which typically has a line width of < 2 nm at RT.[12] However, the higher FWHM in our case indicates these emissions can arise from ensembles of SiV⁻ centers embedded in high quality diamond single crystals.[32] These signatures obtained from the line widths of the color centers clearly establish the perfect crystal quality of the synthesized NDs through thermal oxidation. Other significant observation from these measurements are described further.

The spectral line shape of FL emission also signifies the nature of color centers and interactions with their environment. The spectral line broadening occurs due to the fluctuations in optical resonant frequency either by spectral diffusion or dephasing which decides the line shape as Gaussian or Lorentzian respectively.[43] It is noted that the best fit for the SiV⁻ emission of the NDs in ROI-1, 2 & 3 follows Lorentian profile. The observed Lorentzian emission profile indicates that the spectral broadening is dominated by dephasing process and is negligibly affected by spectral diffusion.

In all NDs irrespective of their size, at 80 and 300 K, fluorescence is observed at ~ 738 nm. Since the quantum efficiency of color centers are very high, the FL emission is much intense than from Raman intensity. Hence, a comparison of these intensities can provide a valuable information about the emission characteristics of the color centers in diamond. For this purpose, we normalize SiV⁻ emission spectrum with Raman intensity measured at identical experimental conditions. At RT, the intensity ratio of SiV⁻ to Raman bands ($I_{SVC}/I_R$) is less than 1 in most of the spectra as can be seen from the Table 1. This shows the presence of large surface states that quenches fluorescence intensity of the SiV⁻ at RT. On the other hand, the intensity of fluorescence from SiV¯ and NV¯ get enhanced significantly at 80 K due to the suppression of free charge carriers from surface states.

Yet another important observation in these studies show that no fluorescent emission at RT from NV¯ of NDs in ROI - 1 and 2. However, some of the NDs from ROI-3 are found to emit even at RT. On the other hand, fluorescence is observed at 80 K at least in the ROI-2 and 3. It is reported that, the fluorescence emission from NV¯ is highly size dependent and rarely observed in particles < 40 nm size. [10,15] This is due migration of vacancy concentrations to surface before they could form vacancy – impurity complex. [9,15] In the present study, the synthesized NDs are having large size distribution and hence, the statistical probability of a ND with size > 50 nm could have been encompassed by a larger laser probe ~ 1 micron. Thus, herein the size is not an issue for the absence of FL emission from NV¯ centers, rather the actual emission could have been possibly quenched due to the effect of laser induced heat and photo bleaching of the NV¯ color centers in the NDs.[15,34]

**Summary**

In summary, a homogeneous distribution of discrete nanodiamonds (NDs) with lateral dimension varying from a few tens of nanometer to about a micron is achieved by controlled thermal oxidation of pre-synthesized nanocrystalline diamond film. The full width at half maximum of the diamond Raman band increases from 2.2 ± 0.5 to 2.8 ± 0.7 cm$^{-1}$ as the mean particle size increase from about 50 nm to a micron. The structural quality of the NDs is found to be extremely high with their unique optical signatures of Raman band of about 1.5 cm$^{-1}$. The NDs with mean particle size of about 50 nm is structurally stable even under 16 mW laser power, while the NDs undergoes structural degradation with increase in laser power when the mean particle size increases above 50 nm. Further, statistical analysis on spectroscopic studies of the NDs confirms that defect density increases with mean particle size. Especially, the presence of NV¯ and GR1 centers are shown up in the spectrum when the mean particle size is above 300 nm. The presence of high density neutral vacancies and NV¯ centers contribute to higher defect concentration in large sized NDs. Nevertheless, the micron sized diamond particles also display several second order diamond Raman bands including 2674 cm$^{-1}$ which is the overtone of 1332

cm$^{-1}$. The observation of second order Raman bands further demonstrate the superior quality of the diamond particles in spite of high density of point defects. In addition, these NDs exhibit a very narrow and intense emission from SiV¯ centers. These discrete NDs can be easily transferred onto a nanoscale instrument which can be used as quantum photon detector or as a source in nano-photonics and scanning near field optical microscopic applications. Additionally, the structurally high quality discrete NDs gives a handle to investigate a lot more fundamental properties of diamond at nanoscale.

# Supplementary information

**Si and N - vacancy color centers in discrete diamond nanoparticles: Raman and fluorescence spectroscopic studies**

**K. Ganesan,[a,*] P. K. Ajikumar,[a] S. Ilango,[a] G. Mangamma [a] and S. Dhara [a]**

[a] *Surface and Nanoscience Division, Materials Science Group, Indira Gandhi Centre for Atomic Research, HBNI, Kalpakkam - 603102, India.*

*\* Corresponding author:* kganesan@igcar.gov.in

Fig. S1. Raman spectra of nanodiamonds in region of interest – 1 with an average lateral size of ~ 50 nm. All the spectra obtained on different locations have very similar features but for a little variation. This can be attributed to the difference in lateral dimension and structural defects. The spectra are stacked up vertically for clarity without removing background.

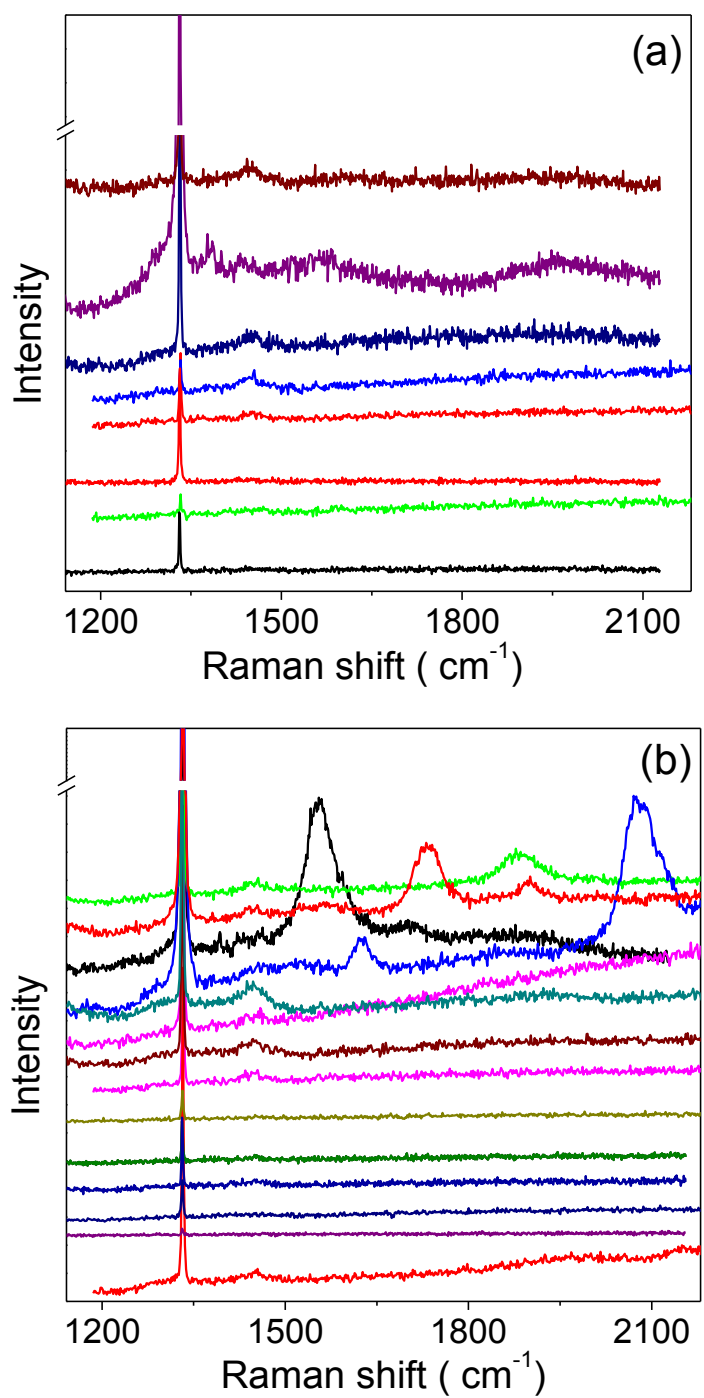

Fig.S2. (a,b) Raman spectra of discrete nanodiamonds in region of interest – 2 with an average lateral size of ~ 300 nm. Each spectrum was recorded on a single diamond nanoparticle. The variation among these spectra indicates that the NDs have structural inhomogeneity that arises due to differences in lateral dimension and structural defects. The spectra are stacked up vertically for clarity without removing background.

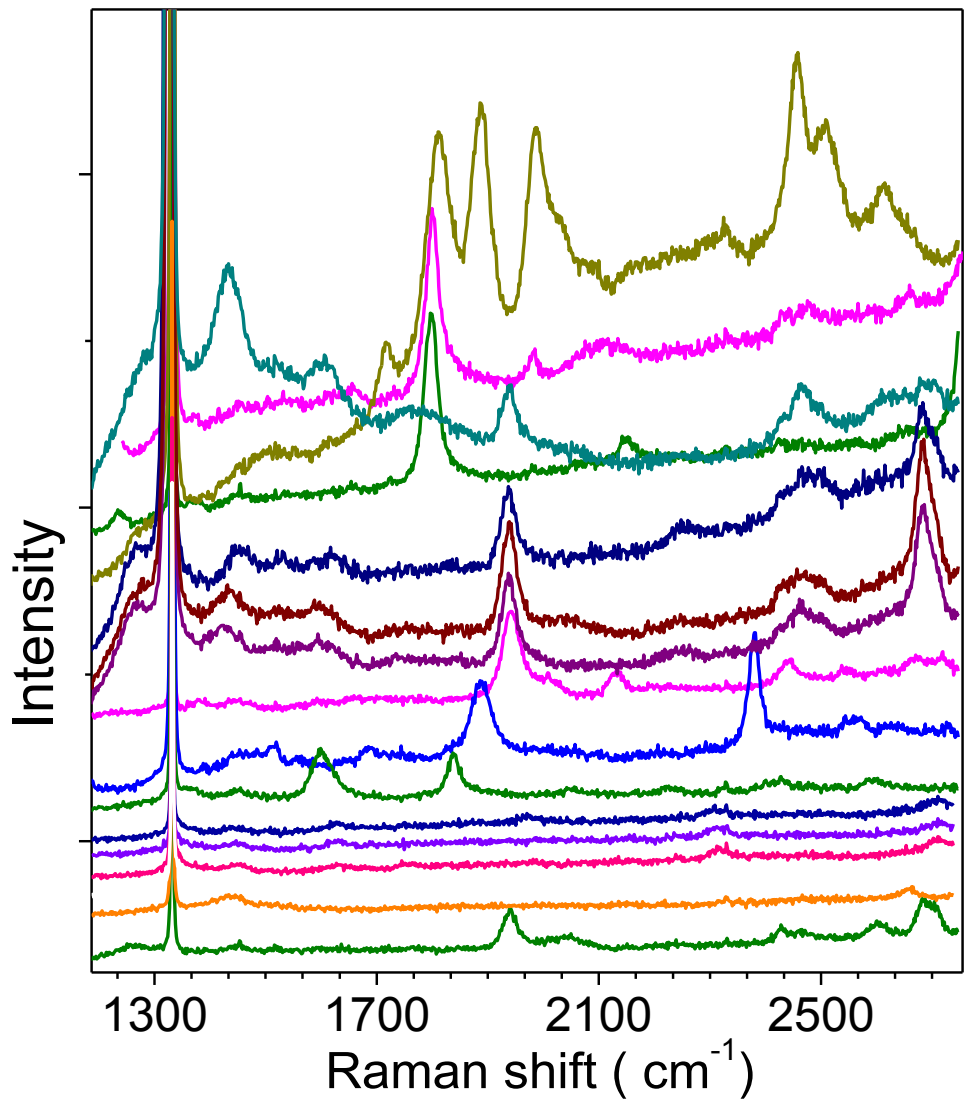

Fig.S3. Raman spectra of discrete diamond particles in region of interest – 3 with an average lateral size of ~ 1 µm. Each spectrum was recorded on a single diamond particle. The variation among these spectra indicates that the particles have structural inhomogeneity that arises due to differences in lateral dimension and structural defects. The spectra are stacked up vertically for clarity without removing background.

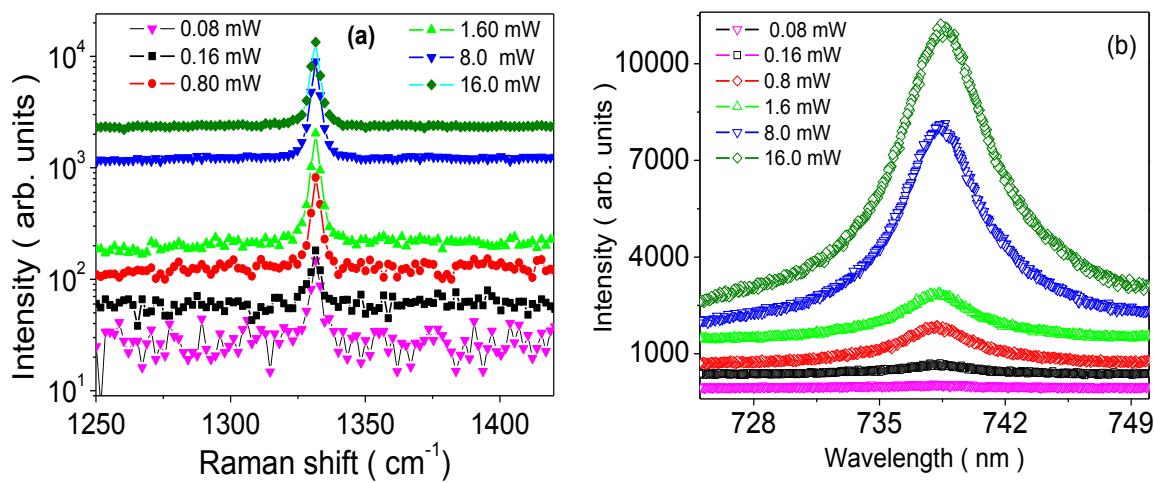

Fig.S4. The intensity variation of (a) Raman and (b) FL spectra as a function of laser power on a discrete diamond nanoparticle in the region of interest – 2. The Raman and FL intensity increases drastically with laser power. Also, a minor change in peak position and line width is observed with increase in laser power. Each spectrum in these plots is recorded for 10 seconds.

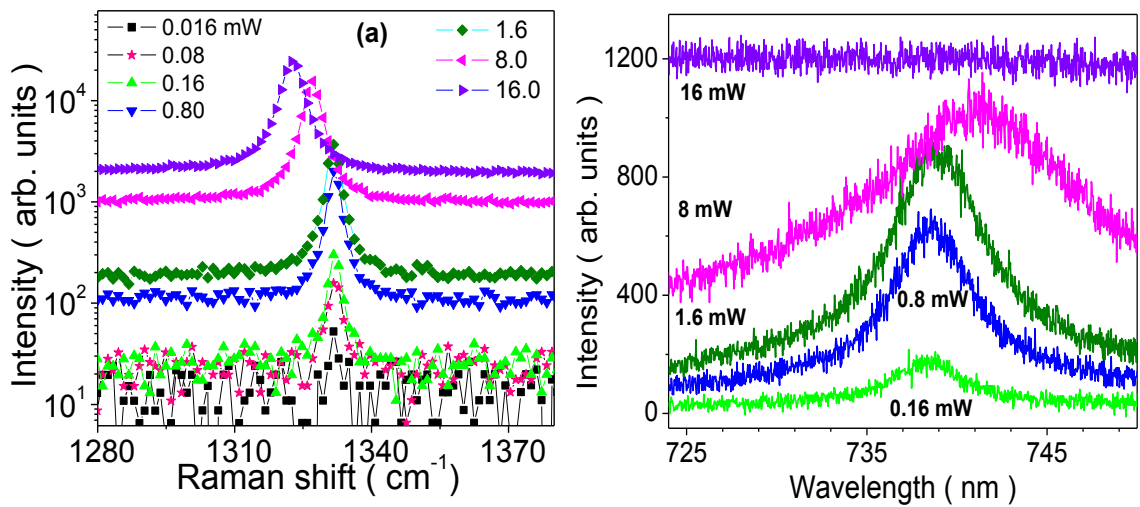

Fig.S5. The intensity variation of (a) Raman and (b) FL spectra as a function of laser power on a discrete diamond nanoparticle in the region of interest – 3. The Raman and FL intensity increases drastically with laser power. Also, a significant change in peak position and line width is observed with increase in laser power. Each spectrum in these plots is recorded for 1 seconds.